\documentclass[conference,10pt]{IEEEtran}
\IEEEoverridecommandlockouts
\usepackage{amsmath,graphicx}
\usepackage{algorithm,algorithmic}
\usepackage{cite,amssymb,amsfonts,color,textcomp}
\usepackage{bm}
\usepackage{balance}
\usepackage{booktabs}
\usepackage{amsthm}
\usepackage{setspace}

\def\thetabf{\boldsymbol \theta}

\def\hbf{{\bf h}}

\def\nbf{{\bf n}}

\def\pbf{{\bf p}}

\def\sbf{{\bf s}}

\def\ubf{{\bf u}}
\def\vbf{{\bf v}}
\def\wbf{{\bf w}}
\def\xbf{{\bf x}}
\def\ybf{{\bf y}}
\def\zbf{{\bf z}}

\def\xbf{{\bf x}}
\def\ybf{{\bf y}}

\def\Hbf{{\bf H}}

\def\Bc{{\cal B}}

\def\Kc{{\cal K}}

\def\Pc{{\cal P}}

\def\Sc{{\cal S}}
\def\Tc{{\cal T}}

\def\Xc{{\cal X}}
\def\Yc{{\cal Y}}

\def\ie{{\it i.e.,\ \/}}
\def\nn{\nonumber}

\def\Re{\mathfrak{R}\mathfrak{e}}
\def\Im{\mathfrak{I}\mathfrak{m}}

\def\mae{{\mathbb{E}}}

\theoremstyle{definition}

\newtheorem{remark}{Remark}
\newtheorem{lemma}{Lemma}

\newtheorem{assumption}{Assumption}
\newtheorem{proposition}{Proposition}

\begin{document}

\title{\huge Joint Downlink-Uplink Beamforming for Wireless Multi-Antenna Federated Learning}

\author{ Chong Zhang$^{\star}$, Min Dong$^{\dagger}$, Ben Liang$^{\star}$, Ali Afana$^{\ddagger}$, Yahia Ahmed$^{\ddagger}$\\
\normalsize $^{\star}$Dept. of Electrical and Computer Engineering, University of Toronto, Canada, $^{\ddagger}$Ericsson Canada, Canada   \\
$^{\dagger}$Dept. of Electrical, Computer and Software Engineering, Ontario Tech University, Canada\thanks{This work was funded in part by Ericsson Canada and by the Natural Sciences and Engineering Research Council (NSERC) of Canada.}
}%

\maketitle

\begin{abstract}
We study joint downlink-uplink beamforming design for wireless federated learning (FL) with a multi-antenna base station. Considering analog transmission over noisy channels and uplink over-the-air aggregation, we derive the global model update expression over communication rounds.
We then obtain an upper bound on the expected global loss function, capturing the downlink and uplink beamforming and receiver noise effect.
We propose a low-complexity joint beamforming algorithm to minimize this upper bound, which employs alternating optimization to breakdown the problem into three subproblems, each solved via closed-form gradient updates. Simulation under practical wireless system setup shows that our proposed joint beamforming design solution substantially outperforms the conventional separate-link design approach and nearly attains the performance of ideal FL with error-free communication links.
\end{abstract}

\section{Introduction}
\label{sec:intro}
Federated learning (FL) \cite{Mcmahan&etal:2017} is a widely recognized machine learning method to process training data locally at multiple worker nodes.
In FL, a parameter server organizes the worker nodes to train a machine learning (ML) model collaboratively using their local datasets.
In the wireless environment, the parameter server can be taken up by a  base station (BS), which exchanges model parameters with participating devices through wireless communication \cite{Zhu&etal:2020}.
However, the fluctuation of the wireless link and noisy reception  bring distortion, leading to  degraded FL performance.
Furthermore, practical wireless systems are limited in  transmission power and bandwidth. This necessitates efficient communication design to effectively support FL, which requires frequent exchange of a massive number of parameters.

Many existing works have considered improving the communication efficiency of FL over wireless channels \cite{Chen&etal:JSAC2012FLWN,Yang&etal:TWC2020,Zhu&etal:TWC2020,Zhang&Tao:TWC2021,Sun&etal:JSAC2022,Fan&etal:TWC2022,Amiri&etal:TWC2022,Wei&Shen}.
Various digital transmission-then-aggregation schemes were proposed for uplink acquisition of local parameters from devices  to the BS \cite{Chen&etal:JSAC2012FLWN}.
Such schemes use conventional  digital transmission via orthogonal channels and can consume a large bandwidth  and incur high latency as the number of devices becomes large.
Later, analog  transmission-and-aggregation schemes were proposed for the uplink
\cite{Yang&etal:TWC2020,Zhu&etal:TWC2020,Zhang&Tao:TWC2021,Sun&etal:JSAC2022,Fan&etal:TWC2022}.
These schemes use analog modulation and superposition  for over-the-air  aggregation of local parameters via the multiple access channel,
substantially saving radio resources over the digital schemes.
However, these works only focused on the uplink, while assuming an error-free downlink.
Subsequently, noisy downlink transmission for FL was studied in \cite{Amiri&etal:TWC2022}
with error-free uplink, where it was shown that since gradient descent in FL is noise resilient, analog transmission can be more efficient than digital transmission, even for the downlink.

In reality, downlink and uplink transmissions are intertwined for parameter exchange in FL. The  quality of one link direction affects the other. Furthermore, the noise and distortion in one communication round propagate to all subsequent communication rounds, which brings challenges to tractable design and analysis. The literature on joint downlink-uplink communication design for FL is scarce.
The convergence of FL with non-i.i.d. local datasets over noisy downlink and uplink channels was recently studied in \cite{Wei&Shen}, where a simple generic signal-in-noise receiver model was used to facilitate analysis without involving actual transmission modeling or design.
Analog design was proposed  for noisy  downlink and uplink
in single-cell \cite{Guo&etal:JSAC2022} or multi-cell \cite{Wang&etal:JSAC2022b} cases.
However, both works considered only single-antenna BSs, and
their solutions and convergence analysis are not applicable to the more practical scenario with a multi-antenna BS.
For multi-antenna communication, transmit and receive beamforming are key techniques to  enhance the communication quality.   While beamforming was considered in  \cite{Amiri&etal:TWC2022} for  FL downlink analog transmission and in   \cite{Yang&etal:TWC2020} for uplink over-the-air aggregation,  there is no existing work on
joint downlink-uplink beamforming design.

In this paper, we study joint downlink-uplink beamforming design to improve the performance of wireless FL with a multi-antenna BS. We consider noisy analog transmission in both directions and uplink over-the-air aggregation for  bandwidth efficiency.
We obtain the overall FL\ global model update over each communication round,  capturing the impact of noisy downlink-uplink transmission and local model updates on FL model training.
Aiming to maximize the training convergence rate, we then derive an upper bound on the expected global loss function after $T$ rounds, and propose a low-complexity joint downlink-uplink beamforming (JDU-BF) algorithm to minimize the upper
bound under  transmit power constraints at the BS and devices. JDU-BF employs the alternating optimization (AO) technique to decompose the joint optimization problem into
three subproblems and solve each via projected gradient descent (PGD) \cite{Levitin&Polyak:USSR1966} with fast closed-form updates. Our simulation results under typical wireless network settings show that
JDU-BF outperforms the conventional separate-link design and provides learning performance close to ideal FL with error-free communication links.

\allowdisplaybreaks
\section{System Model }\label{sec:system_prob}

\subsection{FL System}We consider FL in a wireless network consisting of a server and $K$ worker devices.  Let  $\Kc = \{1, \ldots, K\}$ denote the set of devices. Each device $k \in \Kc$ holds a local training dataset of size $S_k$, denoted by
$\Sc_{k} = \{(\sbf_{k,i},v_{k,i}): 1 \le i \le S_k\}$, where
$\sbf_{k,i}\in\mathbb{R}^{b} $ is the $i$-th data feature vector and $v_{k,i}$ is the label for this data sample. Using their respective local training datasets, the devices  collaboratively
train a global model at the server, represented by the parameter vector $\thetabf\in\mathbb{R}^{D}$,  which predicts the true labels of data feature vectors, while keeping their local datasets private.
The local training loss function that represents the training error
at device $k$ is defined as
\begin{align}
F_{k}(\thetabf)=\frac{1}{S_k}\sum_{i=1}^{S_k} L(\thetabf;\sbf_{k,i},v_{k,i})
\end{align}
where $L(\cdot)$ is the sample-wise training loss associated with each data sample.
 The global training loss function is given by the weighted sum of the local loss functions over all $K$ devices:
\vspace*{-.5em}
\begin{align}
F(\thetabf) = \sum_{k=1}^{K}\frac{S_{k}}{S}F_{k}(\thetabf) \label{eq_global_local_equation}
\end{align}
where $S =  \sum_{k=1}^{K}S_{k}$ is the total number of training samples of all devices.
The learning objective is to find the optimal global model $\thetabf^\star$ that minimizes $F(\thetabf)$.

The devices communicate with the server via noisy downlink and uplink wireless channels to exchange the model update information iteratively for model training. The iterative FL model training procedure in each downlink-uplink communication round $t$ is given as follows:
\begin{itemize}
\item \emph{Downlink broadcast}: The server  broadcasts the current global model parameter vector $\thetabf_t$ to all  $K$ devices via the downlink channels;
\item  \emph{Local model update}: Each device $k$ performs local training independently using its dataset $\Sc_{k}$, based on the received global model $\thetabf_t$. In particular, the device uses the mini-batch approach to divide $\Sc_{k}$ into mini-batches for its local model update, where it performs $J$ iterative local updates and generates the updated local model $\thetabf^{J}_{k,t}$;
\item \emph{Uplink aggregation}: The devices send their updated local models $\{\thetabf^{J}_{k,t}\}_{k\in\Kc}$ to the  server  via the uplink  channels.
The server aggregates $\thetabf^{J}_{k,t}$'s to generate an updated global model $\thetabf_{t+1}$ for the next communication round $t+1$.
\end{itemize}

\subsection{Wireless Communication Model} We consider a practical wireless communication system where the server is
hosted by a BS equipped with $N$ antennas, and each device has a single antenna. The system operates in the time-division duplex (TDD) mode, which is typical for 5G wireless systems. With multiple antennas, the BS uses downlink beamforming to broadcast the global model update $\thetabf_t$ and  applies uplink receiver beamforming to process the received signal from $K$ devices for the global model update.

For the model updating between the BS and devices in the FL system, we consider analog communication for transmitting the updated global/local models. Specifically, the BS and devices   send the respective values of  $\thetabf_t$ and $\{\thetabf^{J}_{k,t}\}_{k\in\Kc}$ directly under their  transmit power budgets.
Furthermore, for the uplink aggregation of the local models, to efficiently use the communication bandwidth, we consider over-the-air computation via analog aggregation over the  multiple access channel. Specifically, the devices send their local model  $\thetabf^{J}_{k,t}$'s  to the BS simultaneously
over the same frequency resources, and  $\thetabf^{J}_{k,t}$'s are aggregated  over the air and received at the BS. Note that the control and signaling channels of the system are still communicated using digital transmissions and are assumed to be perfect.

Due to the noisy communication channels,  the received model updates over  downlink and uplink are  the distorted noisy versions of  $\thetabf_t$ and $\{\thetabf^{J}_{k,t}\}_{k\in\Kc}$, respectively. The errors  in the model updates further propagate over    subsequent communication rounds for FL model training, degrading the learning performance.
 In this paper, we focus on the communication aspect of FL model training. Specifically, our goal is to  jointly design  downlink and uplink beamforming  to maximize the  learning performance of  FL over wireless transmissions.

\section{Downlink-Uplink Transmissions for FL}\label{sec:FL_alg}
We now  formulate
the transmission and reception process with  downlink and uplink beamforming for the FL model update in one communication round.
As mentioned in Section~\ref{sec:system_prob}, each communication round involves three steps. We present each step in detail below.

\subsection{Downlink Broadcast}\label{subsec:dl_broadcast}
At the start of round $t$, the BS has the  current global model,  denoted by
$\thetabf_t = [\theta_{1,t}, \ldots, \theta_{D,t}]^T$.
For efficient transmission, we  convert $\thetabf_t$ into a complex signal vector, whose real and imaginary parts   contain half of the elements in $\thetabf_t$. Specifically,
we can re-express $\thetabf_t = [(\tilde{\thetabf}_t^{\text{re}})^{T}, (\tilde{\thetabf}_t^{\text{im}})^T]^T$, where
$\tilde{\thetabf}_t^{\text{re}} \triangleq [\theta_{1,t}, \ldots, \theta_{\frac{D}{2},t}]^T$, and $\tilde{\thetabf}_t^{\text{im}} \triangleq [\theta_{\frac{D}{2}+1,t}, \ldots, \theta_{D,t}]^T$. Let  $\tilde{\thetabf}_t$ denote the equivalent complex vector representation of $\thetabf_t$, which is given by
$\tilde{\thetabf}_t = \tilde{\thetabf}^{\text{re}}_t + j\tilde{\thetabf}^{\text{im}}_t\in \mathbb{C}^{\frac{D}{2}}$.

  For a TDD system, channel reciprocity holds for downlink and uplink channels. Thus, let $\hbf_{k,t}\in\mathbb{C}^{N}$ denote the  channel vector between the BS
and device $k\in\Kc$ for both downlink and uplink transmissions in round $t$.
We assume $\{\hbf_{k,t}\}_{k\in\Kc}$ remain unchanged during round $t$ and are known perfectly at the BS and the respective devices.
The BS sends the complex global model parameter vector $\tilde{\thetabf}_t$ to the $K$ devices via multicast beamforming.
At round $t$, the received signal vector at device $k$ is given by
\begin{align}
\ubf_{k,t} = (\wbf^{\text{dl}}_t)^H\hbf_{k,t}\tilde{\thetabf}_t + \nbf^{\text{dl}}_{k,t}  \nn
\end{align}
where $\wbf^{\text{dl}}_t\in\mathbb{C}^{N}$  is the downlink multicast beamforming vector at round $t$,
$\nbf^{\text{dl}}_{k,t}\in\mathbb{C}^{\frac{D}{2}}$ is  the receiver additive white Gaussian noise (AWGN)
vector with i.i.d. elements that are zero mean with variance $\sigma^2_{\text{d}}$.
The beamforming vector is subject to the BS transmit power budget. Let  $DP^{\text{dl}}$ be the transmit power budget at the BS for sending the global model $\tilde{\thetabf}_t$ in $D$ channel uses, where $P^{\text{dl}}$ denotes the average transmit power limit per channel use. Then,  for transmitting $\tilde{\thetabf}_t$, $\wbf^{\text{dl}}_t$ is subject to the  transmit power constraint $\|\wbf^{\text{dl}}_t\|^2\|\tilde{\thetabf}_t\|^2 \le DP^{\text{dl}}$.
The BS also sends the scaling factor $\frac{\hbf^{H}_{k,t}\wbf^{\text{dl}}_t }{|\hbf^{H}_{k,t}\wbf^{\text{dl}}_t|^2}$ to device $k$ via the downlink signaling channel to facilitate the receiver processing.
Device $k$  post-processes the received signal $\ubf_{k,t}$ using  the received scaling factor and obtains\begin{align}
\hat{\tilde{\thetabf}}_{k,t} & = \frac{\hbf^{H}_{k,t}\wbf^{\text{dl}}_t }{|\hbf^{H}_{k,t}\wbf^{\text{dl}}_t|^2}\ubf_{k,t}  = \tilde{\thetabf}_t + \tilde{\nbf}^{\text{dl}}_{k,t}
\label{dl_device_signal}
\end{align}
where $\tilde{\nbf}^{\text{dl}}_{k,t} \triangleq \frac{\hbf^{H}_{k,t}\wbf^{\text{dl}}_t }{|\hbf^{H}_{k,t}\wbf^{\text{dl}}_t|^2}\nbf^{\text{dl}}_{k,t}$
is the post-processed noise vector at device $k$.
By the equivalence of real and complex signal representations between  $\thetabf_t$ and $\tilde{\thetabf}_t$,
device $k$ obtains the estimate of the global model $\thetabf_t$,
denoted by $\hat{\thetabf}_{k,t}$,
given by
\begin{align}
\hat{\thetabf}_{k,t} & = \big[\Re{\big\{\hat{\tilde{\thetabf}}_{k,t}\big\}^T}, \Im{\big\{\hat{\tilde{\thetabf}}_{k,t}\big\}^T}\big]^T  = \thetabf_t + \hat{\nbf}^{\text{dl}}_{k,t}
\label{eq_dl}
\end{align}
where $\hat{\nbf}^{\text{dl}}_{k,t} \triangleq [\Re{\{\tilde{\nbf}^{\text{dl}}_{k,t}\}^T},\, \Im{\{\tilde{\nbf}^{\text{dl}}_{k,t}\}^T}]^T$.

\subsection{Local Model Update}\label{subsec:device_update}
Based on  $\hat{\thetabf}_{k,t}$ in \eqref{eq_dl}, device $k$ performs local model training. We assume each device adopts the mini-batch stochastic gradient descent (SGD) algorithm to minimize the local training loss function $F_{k}(\thetabf)$  \cite{Bubeck:2015convex}.
The mini-batch SGD is  a widely adopted training method for ML tasks. It uses a subset of the training dataset
to compute the gradient update at each iteration
and  achieves a favorable tradeoff between computational efficiency and convergence rate.
In particular,  assume that each device applies $J$ mini-batch SGD iterations for its local model update
in each communication round.
Let $\thetabf^{\tau}_{k,t} $ be the local model update by device $k$ at iteration $\tau \in \{0,\ldots,J-1\}$, with
$\thetabf^{0}_{k,t} = \hat{\thetabf}_{k,t}$, and let  $\Bc^{\tau}_{k,t}$ denote the mini-batch, \ie a subset of  $\Sc_{k}$,  at iteration  $\tau+1$.
Then, the local model update is given by
\begin{align}
\thetabf^{\tau+1}_{k,t} & = \thetabf^{\tau}_{k,t} - \eta_t \nabla F_{k}(\thetabf^{\tau}_{k,t}; \Bc^{\tau}_{k,t}) \nn\\
& = \thetabf^{\tau}_{k,t} - \frac{\eta_t}{|\Bc^{\tau}_{k,t}|}\sum_{(\sbf,v)\in\Bc^{\tau}_{k,t}}\!\!\!\nabla L(\thetabf^{\tau}_{k,t}; \sbf,v)
\label{eq_sgd} \\[-2em]\nn
\end{align}
where $\eta_{t}$ is the  learning rate at communication round $t$,
and $\nabla F_{k}$ and $\nabla L$ are the gradient functions  w.r.t. $\thetabf^{\tau}_{k,t}$.
After  $J$ iterations, device $k$ obtains the updated local model $\thetabf^{J}_{k,t}$.
\subsection{Uplink Aggregation}\label{subsec:ul_aggre}
The devices send their updated local models $\{\thetabf^{J}_{k,t}\}_{k\in\Kc}$ to
the BS over their uplink channels and perform over-the-air aggregation.
For efficient transmission similar to
the downlink, we represent $\thetabf^{J}_{k,t}$ using a complex vector,
with the real and imaginary parts of the vector containing the first and second half of the elements in $\thetabf^{J}_{k,t}$, respectively.
Specifically,
we re-write $\thetabf^{J}_{k,t} = [(\tilde{\thetabf}^{J,\text{re}}_{k,t})^T, \, (\tilde{\thetabf}^{J,\text{im}}_{k,t})^T]^T$,
where $\tilde{\thetabf}^{J,\text{re}}_{k,t}\triangleq[\theta^{J}_{k1,t}, \ldots, \theta^{J}_{k\frac{D}{2},t}]^T$
and $\tilde{\thetabf}^{J,\text{im}}_{k,t}\triangleq[\theta^{J}_{k(\frac{D}{2}+1),t}, \ldots, \theta^{J}_{kD,t}]^T$.Then, we have the equivalent complex vector representation of $\thetabf^{J}_{k,t}$, defined by    $\tilde{\thetabf}^{J}_{k,t} = \tilde{\thetabf}^{J,\text{re}}_{k,t} + j\tilde{\thetabf}^{J,\text{im}}_{k,t}\in \mathbb{C}^{\frac{D}{2}}$.

Transmitting  $\tilde{\thetabf}^{J}_{k,t}$ from device $k$ to the BS over its uplink channel requires  $\frac{D}{2}$ channel uses, one for each element in $\tilde{\thetabf}^{J}_{k,t}$. At channel use $l$, the received signal vector at the BS, denoted by $\vbf_{l,t}$, is given by\\[-1.5em]
\begin{align}
\vbf_{l,t} = \sum_{k=1}^{K}\hbf_{k,t}a_{k,t}\tilde{\theta}^{J}_{kl,t} + \ubf^{\text{ul}}_{l,t} \nn
\end{align}
where $a_{k,t}\in\mathbb{C}$ is the  transmit beamforming weight at device $k$, and $\ubf^{\text{ul}}_{l,t}\in\mathbb{C}^N$ is the receiver AWGN vector with i.i.d. elements  that are zero mean
with variance $\sigma^2_{\text{u}}$. The BS applies receive beamforming to  the received signal $\vbf_{l,t}$ over $N$ antennas, for $l=1,\ldots,\frac{D}{2}$, to obtain a weighted sum of $\tilde{\thetabf}^{J}_{k,t}$'s from all $k\in\Kc$. Let $\wbf^{\text{ul}}_t\in\mathbb{C}^{N}$ be the unit-norm receive beamforming vector at the BS at round $t$, with $\|\wbf^{\text{ul}}_t\|^2 = 1$. The post-processed received signal vector over all $\frac{D}{2}$ channel uses is given by
\vspace*{-1em}
\begin{align}
\zbf_t = \sum_{k=1}^{K}(\wbf^{\text{ul}}_t)^H\hbf_{k,t}a_{k,t}\tilde{\thetabf}^{J}_{k,t} + \nbf^{\text{ul}}_t \label{eq_beam_ul} \\[-1.8em] \nn
\end{align}
where $\nbf^{\text{ul}}_t\in\mathbb{C}^{\frac{D}{2}}$ is the post-processed receiver noise with  the $l$-th element being $(\wbf^{\text{ul}}_t)^H\ubf^{\text{ul}}_{l,t}$, for  $l=1,\ldots,\frac{D}{2}$. Define $\alpha^{\text{ul}}_{k,t} \triangleq (\wbf^{\text{ul}}_t)^H\hbf_{k,t}a_{k,t}$, which represents the effective  channel
from device $k$ to the BS after applying  transmit and receive beamforming.
Following this, we re-write \eqref{eq_beam_ul} as
\vspace*{-.5em}
\begin{align}
\zbf_t = \sum_{k=1}^{K}\alpha^{\text{ul}}_{k,t}\tilde{\thetabf}^{J}_{k,t} + \nbf^{\text{ul}}_t.
\label{eq_beam_ul_overall}
\end{align}
\vspace*{-.5em}

We consider  uplink joint transmit and receive beamforming, where $\{a_{k,t}\}_{k\in\Kc}$ and $\wbf^{\text{ul}}_t$ are designed jointly. For over-the-air aggregation, the local models $\tilde{\thetabf}^{J}_{k,t}$'s need to be added up coherently. Thus, the transmit and receive beamforming design should ensure that the resulting effective channels, $\alpha^{\text{ul}}_{k,t}$'s, are phase aligned.
For this purpose, the transmit beamforming weight at device $k$ is set to $a_{k,t} =\sqrt{p_{k,t}}  \frac{\hbf_{k,t}^{H}\wbf^{\text{ul}}_t}{|\hbf^{H}_{k,t}\wbf^{\text{ul}}_t|}$, where $p_{k,t}$ is the transmit power scaling factor for device $k$ at round $t$. Following this, the effective channels of all devices are phase aligned to $0$ after receive beamforming, \ie $\alpha^{\text{ul}}_{k,t}$ is real-valued:
\begin{align*}
\alpha^{\text{ul}}_{k,t} = (\wbf^{\text{ul}}_t)^H\hbf_{k,t}a_{k,t} = \sqrt{p_{k,t}}|\hbf^{H}_{k,t}\wbf^{\text{ul}}_t|, \ k \in \Kc.
\end{align*}
Furthermore, each device is subject to transmit power budget. Let  $DP^{\text{ul}}_k$ be the transmit power budget at device $k$ for sending each local model in $D$ channel uses, where $P^{\text{ul}}_k$ denotes the average transmit power budget per channel use. Then, for transmitting $\tilde{\thetabf}^{J}_{k,t}$, we have the transmit power constraint $p_{k,t}\|\tilde{\thetabf}^{J}_{k,t}\|^2 \le DP^{\text{ul}}_k$.

At the BS receiver, after receive beamforming, the BS further scales $\zbf_t$  in \eqref{eq_beam_ul_overall} to obtain the complex equivalent global model update for the next round $t+1$:\\[-1.5em]
\begin{align}
\tilde{\thetabf}_{t+1} = \frac{\zbf_t}{\sum_{k=1}^{K}\alpha^{\text{ul}}_{k,t}} = \sum_{k=1}^{K}\rho_{k,t}\tilde{\thetabf}^{J}_{k,t} + \tilde{\nbf}^{\text{ul}}_t \label{eq_global_update_0}
\end{align}
where $\rho_{k,t}\!  \triangleq\! \frac{\alpha^{\text{ul}}_{k,t}}{\sum_{j=1}^{K}\alpha^{\text{ul}}_{j,t}}$,  $\sum_{k=1}^{K}\!\rho_{k,t}\! =\! 1$,
 and $\tilde{\nbf}^{\text{ul}}_t \!  \triangleq\! \frac{\nbf^{\text{ul}}_t}{\sum_{k=1}^{K}\alpha^{\text{ul}}_{k,t}}$.

From the local model update in Section~\ref{subsec:device_update}, let
$\Delta\tilde{\thetabf}_{k,t} = \tilde{\thetabf}^{J}_{k,t} - \tilde{\thetabf}^{0}_{k,t}$ denote the  equivalent complex representation of the
local model  change after the local training at device $k$ in round $t$. Based on this and \eqref{dl_device_signal}, we can express the global model $\tilde{\thetabf}_{t+1}$ in \eqref{eq_global_update_0}   in terms of $\tilde{\thetabf}_{t}$ in round $t$ as
\begin{align}
\tilde{\thetabf}_{t+1} &   = \tilde{\thetabf}_t + \sum_{k=1}^{K}\rho_{k,t}\Delta\tilde{\thetabf}_{k,t}
+ \sum_{k=1}^{K}\rho_{k,t}\tilde{\nbf}^{\text{dl}}_{k,t} +  \tilde{\nbf}^{\text{ul}}_t.
\label{eq_global_update}
\end{align}
Finally, the real-valued  global model update $\thetabf_{t+1}$ for round $t+1$ is recovered from $\tilde{\thetabf}_{t+1}$ as
$\thetabf_{t+1}   =  [\Re{\{\tilde{\thetabf}_{t+1}\}^T},\, \Im{\{\tilde{\thetabf}_{t+1}\}^T}]^T$.
\begin{remark}
 The global model updating equation in \eqref{eq_global_update}  is derived from the entire round-trip FL procedure, including downlink-uplink transmission and the local model update at devices.  The second term represents the   aggregated update  from the   local training  at $K$ devices obtained via uplink transmission. The third and fourth noise terms reflect how the  noisy downlink and uplink transmissions affect the global model update.
Overall, the updating equation \eqref{eq_global_update}  shows how local model updates contribute to the global model update under the noisy communication channel and transmitter and receiver processing.
\end{remark}

\section{Joint Downlink-Uplink Beamforming Design}\label{sec:convrg_BF_alg}
In this paper, we consider the design of the communication aspect of the FL system, aiming to maximize the training convergence rate. In particular, we  jointly design the downlink and uplink beamforming  to  minimize the expected global loss function after $T$ rounds. Let $\Tc =\{0,\ldots,T-1\}$.
The optimization problem is formulated as
\begin{align}
\Pc_{o}: &\min_{\{\wbf^{\text{dl}}_t, \wbf^{\text{ul}}_t, \pbf_t\}_{t\in\Tc}} \ \mathbb{E}[F(\thetabf_T)]  \label{obj_expected_gap}\\
\text{s.t.}& \quad \|\wbf^{\text{dl}}_t\|^{2}\|\thetabf_t\|^2 \leq D P^{\text{dl}}, \quad t\in \Tc, \label{constra_dl_T_timeslots} \\
&\quad  p_{k,t}\|\thetabf^{J}_{k,t}\|^2 \leq  D P^{\text{ul}}_k, \ k\in \Kc,t\in \Tc, \label{constra_power_T_timeslots}\\
&\quad  \|\wbf^{\text{ul}}_t\|^2 = 1, \ t\in \Tc \label{constra_ul_T_timeslots}
\end{align}
where $\mathbb{E}[\cdot]$ is the expectation taken w.r.t. receiver noise and mini-batch sampling in local training at each device, and $\pbf_t \triangleq [p_{1,t}, \ldots, p_{K,t}]^T$ contains the uplink transmit power scaling factors of all $K$ devices at round $t$. Constraints in  \eqref{constra_dl_T_timeslots} and \eqref{constra_power_T_timeslots} are the transmit power constraints at the BS and each device $k$, respectively, and constraint in~\eqref{constra_ul_T_timeslots} specifies the receive beamforming vector at the BS to be unit-norm.

Problem $\Pc_{o}$ is  a finite-horizon stochastic optimization problem, which is challenging to solve.
To tackle this problem, we develop a  more tractable upper bound on  $\mathbb{E}[F(\thetabf_T)]$ by analyzing the convergence rate of the global loss function,
 and then we develop a joint downlink-uplink beamforming algorithm to minimize this upper bound.

\subsection{Convergence Analysis on Global Training Loss}\label{subsec:convrg_bound}
Let $F^{\star}$ denote the minimum global loss under the optimal model $\thetabf^{\star}$. To  examine the expected global loss function $\mathbb{E}[F(\thetabf_T)]$ in the  FL system described in Section~\ref{sec:system_prob}, we can equivalently analyze   $\mathbb{E}[F(\thetabf_T)]-F^{\star}$, \ie the expected gap of the global loss function at round $T$ to the minimum global loss, based on the global model updates $\{\thetabf_t\}$ obtained in Section~\ref{sec:FL_alg}.  We first make the following three assumptions on the local loss functions, the SGD, and the difference between the global and weighted average of the local loss functions. These assumptions  are commonly
adopted for the convergence analysis of the FL model training \cite{Amiri&etal:TWC2022,Guo&etal:JSAC2022,Wang&etal:JSAC2022b}.
\vspace*{-.4em}
\begin{assumption}\label{assump_smooth}
The local loss functions $F_k(\cdot)$'s are differentiable and are  $L$-smooth:
$F_k(\ybf) \leq F_k(\xbf) + (\ybf - \xbf)^{T}\nabla F_k(\xbf) + \frac{L}{2}\|\ybf - \xbf\|^2$, $\forall~k\in\Kc$,
$\forall~\xbf, \ybf\in\mathbb{R}^D$.
Also, $F_k(\cdot)$'s are  $\lambda$-strongly convex:
$F_k(\ybf) \geq F_k(\xbf) + (\ybf - \xbf)^{T}\nabla F_k(\xbf) + \frac{\lambda}{2}\|\ybf - \xbf\|^2$, $\forall~k\in\Kc$,
$\forall~\xbf, \ybf\in\mathbb{R}^D$.
\end{assumption}
\vspace*{-1em}

\begin{assumption}\label{assump_unbias}
The mini-batch SGD is unbiased: $\mathbb{E}_{\Bc}[\nabla F_{k}(\thetabf^{\tau}_{k,t}; \Bc^{\tau}_{k,t})] = \nabla F_{k}(\thetabf^{\tau}_{k,t})$, $\forall~k\in\Kc$, $\forall~\tau, t$.
The variance of the mini-batch stochastic gradient is bounded by $\mu$:
For $\forall~k\in\Kc$, $\forall~\tau, t$,
$\mathbb{E}[\|\nabla F_{k}(\thetabf^{\tau}_{k,t}; \Bc^{\tau}_{k,t}) - \nabla F_{k}(\thetabf^{\tau}_{k,t})\|^2] \leq \mu$.
\end{assumption}
\vspace*{-1em}

\begin{assumption}\label{assump_bound_diverg}
The gradient divergence is bounded by $\delta$:
For $\forall~k\in\Kc$, $\forall~\tau, t$,
$\mae[\| \nabla F(\thetabf_t) - \sum_{k=1}^{K}\phi_{k}\nabla F_{k}(\thetabf_t)  \|^2] \leq \delta$,
where $\phi_{k}\in\mathbb{R}$, $\phi_{k}\geq 0$, and $\sum_{k=1}^{K}\phi_{k} = 1$.
\end{assumption}

We now evaluate the expected gap $\mathbb{E}[F(\thetabf_T)]-F^{\star}$ through the training loss convergence rate analysis.
We point out that although  different bounds on this expected gap have been derived in the literature, they are based on either idealized or simplified communication link models without  multi-antenna processing effects. Based on the global model update obtained in \eqref{eq_global_update}, we first bound the expected change of the global loss function in two consecutive rounds as
\vspace*{-.5em}
\begin{align}
\!\!& \mae[F(\thetabf_{t+1})-F(\thetabf_t)]  = \sum_{k=1}^{K}\frac{S_{k}}{S}\mae[F_k(\thetabf_{t+1})-F_k(\thetabf_t)] \nn\\
\!\!& \stackrel{(a)}{\leq}  \mae[(\thetabf_{t+1} - \thetabf_t)^T \nabla F(\thetabf_t)] + \frac{L}{2}\mae[\|\thetabf_{t+1} - \thetabf_t\|^2]  \label{eq_theta_diff_real}\\
\!\!& \stackrel{(b)}{=} \Re\{\mae[(\tilde{\thetabf}_{t+1}-\tilde{\thetabf}_t)^H \nabla\tilde{F}(\thetabf_t)]\}+\frac{L}{2}\mae[\|\tilde{\thetabf}_{t+1}-\tilde{\thetabf}_t\|^2]\!
\nn \\
\!\!& \stackrel{(c)}{=}\!\underbrace{\Re\bigg\{\mae\bigg[\bigg(\sum_{k=1}^{K}\rho_{k,t}\Delta\tilde{\thetabf}_{k,t}
\!+\!\sum_{k=1}^{K}\rho_{k,t}\tilde{\nbf}^{\text{dl}}_{k,t}\!+\!\tilde{\nbf}^{\text{ul}}_t\bigg)^H\nabla \tilde{F}(\thetabf_t)\bigg]\bigg\}}_{\triangleq A_{1,t}}   \nn\\
&\quad+\frac{L}{2}\underbrace{\mae\bigg[\bigg\|\sum_{k=1}^{K}\rho_{k,t}\Delta\tilde{\thetabf}_{k,t}
\!+\!\sum_{k=1}^{K}\rho_{k,t}\tilde{\nbf}^{\text{dl}}_{k,t}\!+\!\tilde{\nbf}^{\text{ul}}_t\bigg\|^2\bigg]}_{\triangleq A_{2,t}}. \label{eq_complex_theta_diff}
\end{align}
where
$(a)$ follows the $L$-smoothness of $F_k(\cdot)$'s in Assumption~\ref{assump_smooth} and the fact that $\nabla F(\thetabf) = \sum_{k=1}^{K}\frac{S_{k}}{S}\nabla F_{k}(\thetabf)$ following \eqref{eq_global_local_equation}, $(b)$ is   the equivalent expression of \eqref{eq_theta_diff_real} by using the equivalent  complex representation    $\tilde{\thetabf}_t$ of $\thetabf_t$, where    $\nabla \tilde{F}(\thetabf_t)$ denotes  the equivalent complex representation of the global loss gradient $\nabla F(\thetabf_t)$ in round $t$, and $(c)$ is obtained following the global model update in \eqref{eq_global_update}.
The upper bound in \eqref{eq_complex_theta_diff} clearly shows the effects of noisy channels and
multi-antenna transmit/receive beamforming processing at both downlink and uplink on the loss function.
Note that $ A_{1,t}$ and $ A_{2,t}$ defined in \eqref{eq_complex_theta_diff} are functions of the aggregated local model change, the downlink-uplink transmission processing, and the receiver noise at round $t$. Next, we  bound $ A_{1,t}$ and $ A_{2,t}$ separately.

For  $ A_{1,t}$, since the receiver noise at the devices and the BS are zero mean and independent of   $\nabla \tilde{F}(\thetabf_t)$, we have
\begin{align}
 A_{1,t}  & =\Re\bigg\{\mae\Big[\Big(\sum_{k=1}^{K}\rho_{k,t}\Delta\tilde{\thetabf}_{k,t}
\Big)^H\nabla \tilde{F}(\thetabf_t)\Big]\bigg\} \label{eq_A1_c}\\[-.5em]
&= \mae\Big[\Big(\sum_{k=1}^{K}\rho_{k,t}\Delta\thetabf_{k,t}
\Big)^T\nabla F(\thetabf_t)\Big] \label{eq_A1}
\end{align}
where  $\Delta\thetabf_{k,t}\triangleq\thetabf^{J}_{k,t} - \thetabf^{0}_{k,t}$ is the real-valued local model  change after the local training at device $k$ in round $t$, and   \eqref{eq_A1} is the equivalent expression of \eqref{eq_A1_c} by using the real-value parameters.  Based on the mini-batch SGD from \eqref{eq_sgd}, we have
$\Delta\thetabf_{k,t} = -\eta_t\sum_{\tau=0}^{J-1} \nabla F_{k}(\thetabf^{\tau}_{k,t}; \Bc^{\tau}_{k,t})$.

Based on Assumptions~\ref{assump_smooth}--\ref{assump_bound_diverg}, we provide an upper bound on $A_{1,t}$, which is stated in the following lemma. Detailed proof is omitted.
Part of our proof has used some techniques in  \cite[Th. 1]{Guo&etal:JSAC2022} (similarly for the proof of Lemma~\ref{lemma2}).
\vspace*{-.5em}
\begin{lemma}\label{lemma1}
 Consider the FL\ system described in Section~\ref{sec:FL_alg} and   Assumptions~\ref{assump_smooth}--\ref{assump_bound_diverg}. Let $Q_t\triangleq1 - 4\eta_t^2J^2L^2$ and assume   $\eta_tJ<\frac{1}{2L}$, $\forall~t\in\Tc$. Then,    $A_{1,t}$ is upper bounded as
\begin{align}
&\hspace*{-.6em}\, A_{1,t} \leq \eta_tJ\bigg(\frac{2}{Q_t}-\frac{5}{2}\bigg)\mae\left[\|\nabla F(\thetabf_t)\|^2\right] \nn\\
& \hspace*{-.6em}  + \frac{D(1-Q_t)}{4\eta_tJQ_t}\sum_{k=1}^{K}\frac{\rho_{k,t}\sigma^2_{\text{d}}}{|\hbf^H_{k,t}\wbf^{\text{dl}}_t|^2} + \frac{\eta_tJ}{2}\bigg(\frac{\delta+\mu }{Q_t}+\frac{\delta-\mu}{2}\bigg). \label{eq_A1_bound}
\end{align}
\end{lemma}
Note that for the bound in \eqref{eq_A1_bound},   $\eta_t$ and $J$ are parameters set in the SGD  for the local model update at each device, and $L, \mu, \delta$ are parameters specified in Assumptions~\ref{assump_smooth}--\ref{assump_bound_diverg}.

For $A_{2,t}$, since the receiver noise at the BS is zero mean and independent of $\sum_{k=1}^{K}\rho_{k,t}(\Delta\tilde{\thetabf}_{k,t}+\tilde{\nbf}^{\text{dl}}_{k,t})$, we have
\begin{align}
\!\!\! A_{2,t} & =  \mae\bigg[\bigg\|\!\sum_{k=1}^{K}\!\rho_{k,t}(\Delta\tilde{\thetabf}_{k,t}\!+\!\tilde{\nbf}^{\text{dl}}_{k,t})\bigg\|^2\bigg]\!+\!\mae[\|\tilde{\nbf}^{\text{ul}}_t\|^2]\nn\\
\!\!\! & = \mae\bigg[\bigg\|\!\sum_{k=1}^{K}\!\rho_{k,t}(\Delta\tilde{\thetabf}_{k,t}\!+\!\tilde{\nbf}^{\text{dl}}_{k,t})\bigg\|^2\bigg] \!+\! \frac{D\sigma^2_{\text{u}}}{2(\sum^{K}_{k=1}\alpha^{\text{ul}}_{k,t})^2}. \label{eq_A2t_firststep}
\end{align}
Based on Assumptions~\ref{assump_smooth}--\ref{assump_bound_diverg},
we can upper bound $A_{2,t}$ as shown in the following lemma.
Detailed proof is omitted.
\begin{lemma}\label{lemma2}
Consider the FL\ system described in Section~\ref{sec:FL_alg} and   Assumptions~\ref{assump_smooth}--\ref{assump_bound_diverg} and assume  $\eta_tJ<\frac{1}{2L}$, $\forall~t\in\Tc$. Then,
$A_{2,t}$ is upper bounded as
\begin{align}
\!\!\!\!& A_{2,t} \!\leq\! \frac{2}{L^2}\bigg(\frac{1-Q_t}{Q_t}\bigg)\mae[\|\nabla F(\thetabf_t)\|^2] \nn\\
&+D\bigg(\frac{1-Q_t}{Q_t}\sum_{k=1}^{K}\!\!\frac{\rho_{k,t}\sigma^2_{\text{d}}}{|\hbf^H_{k,t}\wbf^{\text{dl}}_t|^2}+\sum_{k=1}^{K}\!\!\frac{\rho_{k,t}^2\sigma^2_{\text{d}}}{|\hbf^H_{k,t}\wbf^{\text{dl}}_t|^2}\bigg) \nn\\
\!\!\!\!&\! +\!\frac{D\sigma^2_{\text{u}}}{2(\sum^{K}_{k=1}\!\alpha^{\text{ul}}_{k,t})^2}\!+\frac{1-Q_t}{2L^{2}Q_t}\bigg(\Big(1-Q_t+\frac{Q_t}{J}\Big)\mu+ 4\delta\bigg). \label{eq_A2t_lemma2}
\end{align}
\end{lemma}

We now  analyze the expected gap $\mathbb{E}[F(\thetabf_T)]-F^{\star}$ at round $T$. From  \eqref{eq_complex_theta_diff},  the expected gap  at round $t+1$ is bounded as
\begin{align}\label{theta_diff}
\mae[F(\thetabf_{t+1})]- F^{\star}\le \mae[F(\thetabf_t)]- F^{\star}+ A_{1,t}+\frac{L}{2}A_{2,t}.
\end{align}
 Using Lemmas~\ref{lemma1} and \ref{lemma2},
 we can further bound the right hand side (RHS) of \eqref{theta_diff}.  Summing up both sides over $t\in\Tc$ and rearranging the terms, we  can obtain the upper bound on $\mathbb{E}[F(\thetabf_T)] - F^{\star}$, which is stated in Proposition~\ref{thm:convergence}
 below.
\begin{proposition}\label{thm:convergence}
For the FL\ system described in Section~\ref{sec:FL_alg},  under  Assumptions~\ref{assump_smooth}--\ref{assump_bound_diverg} and for  $\frac{1}{10L}\le \eta_tJ<\frac{1}{2L}$, $\forall~t\in\Tc$, the expected  gap   $\mathbb{E}[F(\thetabf_T)] - F^{\star}$
after $T$ communication rounds is upper bounded by
\begin{align}
\mathbb{E}[ F(\thetabf_T)] -  F^{\star}\! \leq & \ \Gamma\!\prod_{t=0}^{T-1}\!G_t\! +\Lambda+\! \sum_{t=0}^{T-2}\!H(\wbf^{\text{dl}}_t,  \wbf^{\text{ul}}_t, \pbf_t)\!\!\!\prod_{s=t+1}^{T-1}\!\!G_s \nn\\
& + H(\wbf^{\text{dl}}_{T-1}, \wbf^{\text{ul}}_{T-1}, \pbf_{T-1})
\label{eq_thm1}
\end{align}
where  $\Gamma \triangleq \mathbb{E}[ F(\thetabf_0)] - F^{\star}$, $\Lambda \triangleq \sum_{t=0}^{T-2}C_t \big(\prod_{s=t+1}^{T-1}G_s\big)+ C_{T-1}$ with
\begin{align*}
&G_t \triangleq \frac{1-Q_t}{4\eta_tJ \lambda Q_t}\big(5(1-Q_t)+4\sqrt{1-Q_t}-1\big) + 1,\\
&C_t \triangleq \frac{\eta_tJ}{2}\!\Big(\frac{\delta+\mu }{Q_t}\!+\!\frac{\delta\!-\!\mu}{2}\Big)\!+\!\frac{1\!-\!Q_t}{2L^{2}Q_t}\Big(\Big(1\!-\!Q_t\!+\!\frac{Q_t}{J}\Big)\mu\!+\! 4\delta\Big),
\end{align*}
and $H(\wbf^{\text{dl}}_t, \wbf^{\text{ul}}_t, \pbf_t)$ is defined in \eqref{eq_fc_Ht}.
\end{proposition}
\vspace*{-.3em}
\begin{IEEEproof}
See Appendix~\ref{appC}.
\end{IEEEproof}
\begin{figure*}[!htbp]
\small \begin{align}
H(\wbf^{\text{dl}}_t, \wbf^{\text{ul}}_t, \pbf_t) \triangleq
&  \frac{LD}{2}\!\bigg( \frac{1\!-\!Q_t\!+\!\sqrt{1-Q_t}}{Q_t}\bigg)
\frac{\displaystyle \sigma^{2}_{\text{d}}\left(\sum^{K}_{k=1}\frac{\sqrt{p_{k,t}}|\hbf_{k,t}^H\wbf^{\text{ul}}_t|}{|\hbf_{k,t}^H\wbf^{\text{dl}}_t|^2}\right)}{\displaystyle\sum_{k=1}^{K}\sqrt{p_{k,t}}|\hbf_{k,t}^H\wbf^{\text{ul}}_t|} + \! \frac{LD}{2}\frac{\displaystyle \sigma^{2}_{\text{d}}\left(\sum^{K}_{k=1}\frac{p_{k,t}|\hbf_{k,t}^H\wbf^{\text{ul}}_t|^2}{|\hbf_{k,t}^H\wbf^{\text{dl}}_t|^2}\right)+\frac{\sigma_\text{u}^2}{2}}{\bigg(\displaystyle\sum_{k=1}^{K}\sqrt{p_{k,t}}|\hbf_{k,t}^H\wbf^{\text{ul}}_t|\bigg)^2} \label{eq_fc_Ht} \\[.5em]
\hline \nn
\end{align}
\vspace*{-3em}
\end{figure*}

Note that the upper bound for $\mathbb{E}[F(\thetabf_T)] - F^{\star}$ in  \eqref{eq_thm1}
reflects how the downlink-uplink transmission and the  local training  affect the convergence of
the global model update.
In particular, the first term shows the impact of  the  initial starting point $\thetabf_0$. The second term $\Lambda$ is a weighted sum of $C_t$'s, each accounting for the  gradient divergence of the local loss function from the global loss function using the mini-batch SGD  during the local model updates among $K$ devices in round $t$.\footnote{Recall that $\lambda$ in the expression of $G_t$ is specified in Assumption~\ref{assump_smooth}.} The third  term is a weighted sum of $H(\wbf^{\text{dl}}_t, \wbf^{\text{ul}}_t, \pbf_t)$, where   $H(\wbf^{\text{dl}}_t, \wbf^{\text{ul}}_t, \pbf_t)$ in \eqref{eq_fc_Ht} is in the form of a weighted sum of the inverse of SNRs (\ie noise-to-signal ratio). Two types of SNRs  are shown: the terms with  $\sigma_\text{d}^2$ reflect the post-processing SNR at the BS receiver due to the  downlink device receiver noise  (after  downlink and uplink beamforming and receiver processing), and the term with  $\sigma_\text{u}^2$ shows the post-processing SNR at the BS receiver due to the  BS receiver noise in the uplink after receiver beamforming and processing.

The upper bound given in Proposition~\ref{thm:convergence} is in a  more tractable form  for the expected  gap $\mathbb{E}[F(\thetabf_T)] - F^{\star}$
that we can use for the optimization design.
In the following, we directly minimize this upper bound to
obtain a joint downlink and uplink beamforming solution.

\subsection{Joint Downlink-Uplink Beamforming Algorithm}\label{sec:joint_design}
We now replace the objective function in $\Pc_{o}$   with the upper bound
in \eqref{eq_thm1}.
Define \begin{align}
&\Psi(\{\wbf^{\text{dl}}_t, \wbf^{\text{ul}}_t, \pbf_t\}) \nn\\
 & \triangleq \sum_{t=0}^{T-2}\!H(\wbf^{\text{dl}}_t,  \wbf^{\text{ul}}_t,\pbf_t)\!\!\!\prod_{s=t+1}^{T-1}\!\!G_s  + H(\wbf^{\text{dl}}_{T-1}, \wbf^{\text{ul}}_{T-1}, \pbf_{T-1}). \nn
\end{align}
Omitting the constant terms $\Gamma\prod_{t=0}^{T-1}G_t + \Lambda$ in \eqref{eq_thm1},
we can equivalently  minimize  $\Psi(\{\wbf^{\text{dl}}_t, \wbf^{\text{ul}}_t, \pbf_t\})$. Thus, instead of $\Pc_o$, we consider the following joint  downlink and uplink
beamforming optimization problem
\begin{align*}
\Pc_{1}: \min_{\{\wbf^{\text{dl}}_t, \wbf^{\text{ul}}_t, \pbf_t\}_{t=0}^{T-1}} & \ \ \Psi(\{\wbf^{\text{dl}}_t, \wbf^{\text{ul}}_t, \pbf_t\}) &
\text{s.t.}& \ \ \eqref{constra_dl_T_timeslots}\eqref{constra_power_T_timeslots}\eqref{constra_ul_T_timeslots}.
\end{align*}

Problem $\Pc_{1}$ is a  $T$-horizon joint optimization problem that includes $T$ communication rounds of the model update. Note that in Proposition~\ref{thm:convergence}, for  $\frac{1}{10L}\le \eta_tJ<\frac{1}{2L}$, we have  $G_t > 0$, $\forall~t\in\Tc$, and thus, $\prod_{s=t+1}^{T-1} G_s > 0$ in $\Psi(\{\wbf^{\text{dl}}_t, \wbf^{\text{ul}}_t, \pbf_t\})$.
Thus,  $\Pc_{1}$ can be decomposed into $T$ subproblems,
one for each round $t$ given by
\vspace*{-.5em}
\begin{align}
\Pc_{2}^t: \min_{\wbf^{\text{dl}}_t, \wbf^{\text{ul}}_t, \pbf_t}&  H(\wbf^{\text{dl}}_t, \wbf^{\text{ul}}_t, \pbf_t) \nn\\
\text{s.t.}\ \ &  \|\wbf^{\text{dl}}_t\|^{2}\|\thetabf_t\|^2 \leq D P^{\text{dl}}, \label{constra_dl}\\
&   p_{k,t}\|\thetabf^{J}_{k,t}\|^2 \leq  D P^{\text{ul}}_k, \,\, k\in\Kc, \\
&  \|\wbf^{\text{ul}}_t\|^2 = 1. \label{constra_ul}
\end{align}

Note that  $H(\wbf^{\text{dl}}_t, \wbf^{\text{ul}}_t, \pbf_t)$ in \eqref{eq_thm1} is an involved non-convex function of  $(\wbf^{\text{dl}}_t, \wbf^{\text{ul}}_t, \pbf_t)$. It is difficult to find the optimal solution to  $\Pc_2^t$ directly. Instead, we  propose to use the AO approach to solve $\Pc_2^t$ w.r.t.  the downlink beamforming $\wbf^{\text{dl}}_t$, and uplink beamforming $(\wbf^{\text{ul}}_t, \pbf_t)$ alternatingly. Furthermore,  we propose to solve each AO subproblem via PGD \cite{Levitin&Polyak:USSR1966}.

To facilitate the computation in our algorithm,
we express all the complex quantities in $\Pc_{2}^t$  using their real and imaginary parts.
Define $\xbf^{\text{dl}}_t\triangleq [\Re{\{\wbf^{\text{dl}}_t\}}^T, \Im{\{\wbf^{\text{dl}}_t\}}^T]^T$,
$\xbf^{\text{ul}}_t\triangleq [\Re{\{\wbf^{\text{ul}}_t\}}^T, \Im{\{\wbf^{\text{ul}}_t\}}^T]^T$,
and
\begin{align}
\Hbf_{k,t}\triangleq
\begin{bmatrix}
\Re{\{\hbf_{k,t}\hbf^H_{k,t}\}} & \!\!\! -\Im{\{\hbf_{k,t}\hbf^H_{k,t}\}} \\
\Im{\{\hbf_{k,t}\hbf^H_{k,t}\}} &  \!\!\!\Re{\{\hbf_{k,t}\hbf^H_{k,t}\}} \nn
\end{bmatrix},\ k\in\Kc, t\in\Tc.
\end{align}
Then, we have $\|\wbf^{\text{dl}}_t\|^{2}=\|\xbf^{\text{dl}}_t\|^2$,
$\|\wbf^{\text{ul}}_t\|^{2}=\|\xbf^{\text{ul}}_t\|^2$,
$|\hbf_{k,t}^H\wbf^{\text{dl}}_t|^2=(\xbf^{\text{dl}}_t)^T\Hbf_{k,t}\xbf^{\text{dl}}_t$, and
$|\hbf_{k,t}^H\wbf^{\text{ul}}_t|^2=(\xbf^{\text{ul}}_t)^T\Hbf_{k,t}\xbf^{\text{ul}}_t$.
Thus, we can  express $H(\wbf^{\text{dl}}_t, \wbf^{\text{ul}}_t, \pbf_t)$ in \eqref{eq_fc_Ht}  using  the corresponding real-valued vectors $(\xbf^{\text{dl}}_t, \xbf^{\text{ul}}_t, \pbf_t)$ by replacing $|\hbf_{k,t}^H\wbf^{\text{dl}}_t|$ and $|\hbf_{k,t}^H\wbf^{\text{ul}}_t|$ in  \eqref{eq_fc_Ht} with $((\xbf^{\text{dl}}_t)^T\Hbf_{k,t}\xbf^{\text{dl}}_t)^\frac{1}{2}$ and $((\xbf^{\text{ul}}_t)^T\Hbf_{k,t}\xbf^{\text{ul}}_t)^\frac{1}{2}$, respectively.
Let  $\Phi(\xbf^{\text{dl}}_t, \xbf^{\text{ul}}_t, \pbf_t)$ denote this resulting equivalent converted function from $H(\wbf^{\text{dl}}_t, \wbf^{\text{ul}}_t, \pbf_t)$.  Then,   $\Pc_{2}^t$ can be equivalently transformed into the following problem with all real-valued optimization variables:
\begin{align*}
\Pc_{3}^t: \min_{\xbf^{\text{dl}}_t, \xbf^{\text{ul}}_t, \pbf_t} \ & \Phi(\xbf^{\text{dl}}_t, \xbf^{\text{ul}}_t, \pbf_t)  \nn\\
 \text{s.t.} \ & \xbf^{\text{dl}}_t\in\Xc^{\text{dl}}_t, \xbf^{\text{ul}}_t\in\Xc^{\text{ul}}_t,  \pbf_t\in\Yc_t
\end{align*}
where $\Xc^{\text{dl}}_t\triangleq \{\xbf^{\text{dl}}_t: \|\xbf^{\text{dl}}_t\|^{2}\|\thetabf_t\|^2 \leq D P^{\text{dl}}\}$,
$\Xc^{\text{ul}}_t\triangleq \{\xbf^{\text{ul}}_t: \|\xbf^{\text{ul}}_t\|^2 = 1\}$,
and $\Yc_t\triangleq \{\pbf_t: p_{k,t}\|\thetabf^{J}_{k,t}\|^2 \leq  D P^{\text{ul}}_k, \,\, k\in\Kc\}$.

We   use the AO approach to compute a solution to $\Pc_{3}^t$ at each round $t\in\Tc$.
Our proposed JDU-BF algorithm for $\Pc_1$ is summarized in  Algorithm~\ref{alg:alt_alg}.
Note that  subproblem \eqref{dl_BF_update_AO} is a downlink beamforming problem,  subproblem \eqref{ul_BF_update_AO} is an uplink receive beamforming problem, and subproblem \eqref{power_update_AO} is  an uplink transmit power minimization problem. Thus, our proposed algorithm solves downlink and uplink beamforming problems alternatingly.

\begin{algorithm}[t]
\caption{The JDU-BF Algorithm for $\Pc_{1}$}
\label{alg:alt_alg}
\begin{algorithmic}[0]
\STATE \textbf{Initialization:} Set $t=0$.
\REPEAT
\STATE \textbf{Initialization:} Set $\xbf^{\text{dl}(0)}_{t}, \xbf^{\text{ul}(0)}_{t}, \pbf^{(0)}_{t}$; Set $i = 0$.
\REPEAT
\STATE  1) Update downlink transmit beamforming vector\\[-1.5em]
 \begin{align}
\xbf^{\text{dl}(i+1)}_{t}= & \mathop{\arg\min}\limits_{\xbf^{\text{dl}}_t\in\Xc^{\text{dl}}_t}\Phi(\xbf^{\text{dl}}_t,  \xbf^{\text{ul}(i)}_{t}, \pbf^{(i)}_{t}). \label{dl_BF_update_AO}
\end{align}\vspace*{-1em}
\STATE 2) Update uplink receive beamforming vector\\[-1.5em]
 \begin{align}
\xbf^{\text{ul}(i+1)}_{t}= & \mathop{\arg\min}\limits_{\xbf^{\text{ul}}_t\in\Xc^{\text{ul}}_t}\Phi(\xbf^{\text{dl}(i+1)}_{t}, \xbf^{\text{ul}}_t,  \pbf^{(i)}_{t}). \label{ul_BF_update_AO}
\end{align}\vspace*{-1em}
\STATE 3) Update uplink device transmit power \\[-1.5em]
 \begin{align}
\pbf^{(i+1)}_t= & \mathop{\arg\min}\limits_{\pbf_t\in\Yc_t}\Phi(\xbf^{\text{dl}(i+1)}_t, \xbf^{\text{ul}(i+1)}_t, \pbf_t). \label{power_update_AO}
\end{align}\vspace*{-1em}
\STATE 4) Set $i \leftarrow i+1$.
\UNTIL{convergence // solve $\Pc_{3}^t$}
\STATE Set $t\leftarrow t+1$.
\UNTIL{$t=T$}
\end{algorithmic}
\end{algorithm}

For each subproblem in \eqref{dl_BF_update_AO}--\eqref{power_update_AO}, the objective function is a complicated  non-convex function of the optimization variable. Thus, we adopt PGD to solve each subproblem. PGD  \cite{Levitin&Polyak:USSR1966} is an iterative first-order algorithm that uses gradient updates to solve a  constrained minimization problem:
$\min_{\xbf\in\Xc} f(\xbf)$, where $\Xc$ is the convex feasible set for $\xbf$.   PGD has the following updating procedure: At iteration $j$,
\begin{align}
\xbf_{j+1} = \Pi_{\Xc}\big(\xbf_j - \beta\nabla_{\xbf}f(\xbf_j)\big) \label{PGD_update}
\end{align}
where $\beta > 0$ is the step size and $\Pi_\Xc(\xbf)$ denotes the projection of point $\xbf$ onto set $\Xc$.
Note that due to the inherent structure of our problem,  PGD
is particularly suitable for solving subproblems \eqref{dl_BF_update_AO}--\eqref{power_update_AO} at each AO iteration.
In particular,
the projection $\Pi_\Xc(\xbf)$ operation can be expressed in closed-form for each of subproblems \eqref{dl_BF_update_AO}--\eqref{power_update_AO}:
\begin{itemize}
\item  For subproblem \eqref{dl_BF_update_AO}:\\[-1em]
\begin{align}
 \Pi_{\Xc^{\text{dl}}_t}(\xbf^{\text{dl}}_t)=
\begin{cases}
\sqrt{\frac{D P^{\text{dl}}}{\|\xbf^{\text{dl}}_t\|^{2}\|\thetabf_t\|^2}}\xbf^{\text{dl}}_t  & {\xbf^{\text{dl}}_t\notin\Xc^{\text{dl}}_t}, \\
\xbf^{\text{dl}}_t          & {\xbf^{\text{dl}}_t\in\Xc^{\text{dl}}_t}.
\end{cases}
\nn
\end{align}
\item For subproblem \eqref{ul_BF_update_AO}:
$\Pi_{\Xc^{\text{ul}}_t}(\xbf^{\text{ul}}_t)=\frac{\xbf^{\text{ul}}_t}{\|\xbf^{\text{ul}}_t\|}$.

\item For subproblem \eqref{power_update_AO}:  $\Pi_{\Yc_t}(\pbf_t)$ is given by
\begin{align}
  p_{k,t}=
\begin{cases}
\frac{D P^{\text{ul}}_k}{\|\thetabf^{J}_{k,t}\|^2}    & p_{k,t}>\frac{D P^{\text{ul}}_k}{\|\thetabf^{J}_{k,t}\|^2}, \\
p_{k,t}         & p_{k,t}\leq\frac{D P^{\text{ul}}_k}{\|\thetabf^{J}_{k,t}\|^2}.
\end{cases} \quad \forall~k\in\Kc
\nn
\end{align}
\end{itemize}
Thus, the computation using PGD via \eqref{PGD_update} has low complexity.
Also, PGD is guaranteed to find an approximate stationary point for each subproblem in polynomial time \cite{Ghadimi&etal:Math2016}.
We summarized our proposed JDU-BF algorithm in Algorithm~\ref{alg:alt_alg}.
\vspace*{-1.3em}
\subsection{Separate Downlink and Uplink Beamforming Design}\label{sec:seperate_design}
In the above, we have proposed joint downlink-uplink beamforming design for the FL\ system, which is based on the global model update in \eqref{eq_global_update} derived from each communication round. For comparison purpose,  we also consider the conventional approach where downlink and uplink transmission are designed separately for the communication system.

\emph{Downlink:} We formulate the problem of  downlink beamforming  and the uplink beamforming  separately. At the communication round $t$,  since the BS broadcasts the global model to all devices, the downlink beamforming problem is to maximize the minimum received SNR, which is a single-group multicast beamforming max-min fair problem:
\begin{align}
\max_{\wbf^{\text{dl}}_t}\min_{k\in\Kc} \,\, |\hbf_{k,t}^H\wbf^{\text{dl}}_t|^2 \quad \text{s.t.} \quad \|\wbf^{\text{dl}}_t\|^{2}\|\thetabf_t\|^2 \leq D P^{\text{dl}}. \label{dl_BF_update_SDUBF}
\end{align}
The solution to this problem can be efficiently computed using the projected subgradient algorithm proposed in \cite{Zhang&etal:WCL2022} based on the  optimal multicast beamforming structure \cite{Dong&Wang:TSP2020}.

\emph{Uplink:} For uplink  over-the-air aggregation,   the transmit beamforming weight at device $k$ is set to $a_{k,t} =\sqrt{p_{k,t}}  \frac{\hbf_{k,t}^{H}\wbf^{\text{ul}}_t}{|\hbf^{H}_{k,t}\wbf^{\text{ul}}_t|}$ to phase-align the transmissions from all devices. Each device uses the maximum transmit power; thus, the power scaling factor is $p_{k,t}=\frac{D P^{\text{ul}}_k}{\|\thetabf^{J}_{k,t}\|^2}$, $\forall~ k\in\Kc$. Then, we
 design uplink receive beamforming to maximize the received SNR (from the aggregated signal) at the BS:
\vspace*{-.5em}
\begin{align}
\max_{\wbf^{\text{ul}}_t: \|\wbf^{\text{ul}}_t\|^{2}=1}\sum_{k=1}^{K} p_{k,t}|\hbf_{k,t}^H\wbf^{\text{ul}}_t|^2.   \label{ul_BF_update_SDUBF}
\end{align}
This problem can be solved  using PGD via \eqref{PGD_update}. We name this approach as the separate downlink and uplink beamforming (SDU-BF) algorithm.

\section{Simulation Results}
\label{sec:simulations}

\subsubsection{Simulation Setup}
We consider the real-world dataset for image classification under an LTE wireless system setting. Following the typical LTE specifications,
we set system  bandwidth  $10$~MHz and carrier frequency $2$~GHz.
The maximum BS transmit power is $47~\text{dBm}$. The maximum device transmit power is $23~\text{dBm}$, and we assume the devices use $1$ MHz bandwidth for uplink transmission.
The path gain between the BS and device $k$  is
$G_{k} [\text{dB}] = -139.2-35\log_{10}d_k - \psi_k$,
where $d_k\in(1\text{~km}, 1.5\text{~km})$ is the BS-device distance in kilometers, and  $\psi_k$ is the shadowing random variable with standard deviation  $8~\text{dB}$.
The channel vector is generated as $\hbf_{k,t} = \sqrt{G_{k}}\bar{\hbf}_{k,t}$ with
$\bar{\hbf}_{k,t}\sim\mathcal{CN}({\bf{0}},{\bf{I}})$. Noise power spectral density is $N_0 = -174~\text{dBm/Hz}$,
and noise figure $N_F=8$~dB and $2$ dB at the device and  BS receivers, respectively.

We adopt the MNIST dataset \cite{Lecun&etal:MNIST} for  model training and testing.
MNIST consists of $6\times 10^4$ training samples and $1\times 10^4$ test samples from 10 different classes. Each sample
is a labeled image of size $28\times 28$ pixels,
\ie $\sbf\in\mathbb{R}^{784}$ and $v\in\{0,\ldots,9\}$ indicating the class.
We consider training a convolutional neural network with an $8\times3\times3$ ReLU convolutional layer,
a $2\times2$ max pooling layer, a ReLU fully-connected layer, and a softmax output layer,
resulting in $D=1.361\times 10^4$ model parameters in total.
We use the $1\times 10^4$ test samples to measure the test accuracy of the global model update $\thetabf_t$ at each round $t$. The training samples are randomly and evenly distributed  over devices, and the local dataset at device $k$ has $S_{k} = \frac{6\times 10^4}{K}$ samples.
For the local training via the SGD at each device, we set  $L=10$, $J=30$, mini-batch size $|\Bc^{\tau}_{k,t}|=\frac{2\times 10^3}{K}, \forall k,\tau,t$, and the learning rate $\eta_t=\frac{1}{10JL}$, $\forall t$.
\subsubsection{Performance Comparison}
For the comparison purpose, we consider the following three schemes:
i) \textbf{Ideal FL}\cite{Mcmahan&etal:2017}: Perform FL\ via the global model update in \eqref{eq_global_update}, assuming error-free downlink and uplink and perfect recovery of model parameters at the BS and devices, \ie receiver noise $\tilde{\nbf}^{\text{dl}}_{k,t} =  \tilde{\nbf}^{\text{ul}}_t = \bf{0}$,
 receiver post-processing weight $\rho_{k,t}=\frac{1}{K}$, $\forall k,t$.
This   benchmark provides the performance upper bound for all schemes.
ii) \textbf{SDU-BF}: the separate SNR-maximizing design scheme described in Section~\ref{sec:seperate_design}.
iii) \textbf{Random beamforming (RBF)}: Perform FL via \eqref{eq_global_update} with randomly generated downlink and uplink beamforming vectors $\wbf^{\text{dl}}_t$ and $\wbf^{\text{ul}}_t$. The devices use the maximum transmit power and do not perform transmit beamforming phase alignment.

\begin{figure*}[t]
\centering
\hspace*{-.8em}\includegraphics[scale=.58]{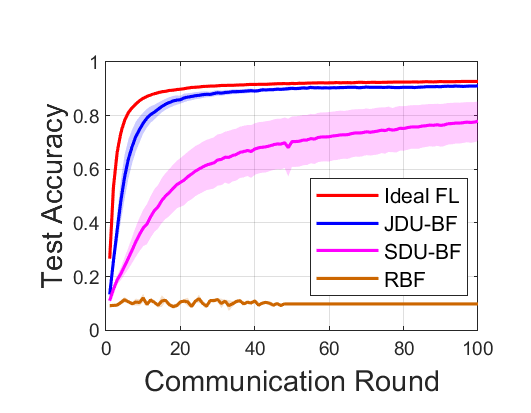}
\hspace*{-0em}\includegraphics[scale=0.58]{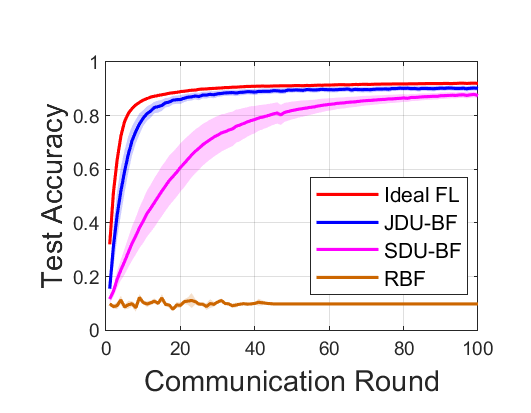}
\hspace*{-0em}\includegraphics[scale=.58]{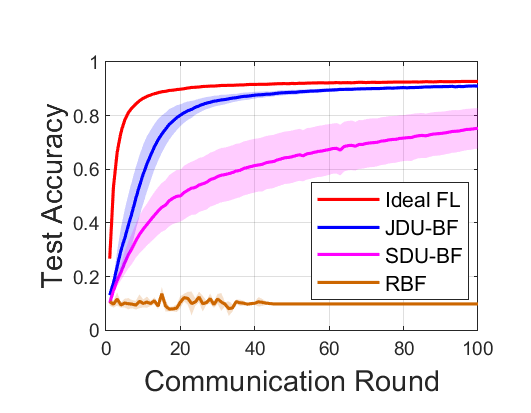}
\vspace*{-1em}\caption{Test accuracy vs. communication round $T$.  Left: $N = 64$,\,$K = 20$. Middle: $N = 64$,\,$K = 40$. Right: $N = 16$,\,$K = 20$.}
\label{Fig1:Accuracy_N16_64_K20_40} \vspace*{-1em}
\end{figure*}

Fig.~\ref{Fig1:Accuracy_N16_64_K20_40} shows the test accuracy performance by the considered methods over communication round $T$ for three system settings  for $(N,K)$.
All curves are obtained by averaging over $20$ channel realizations.
The shadowed area over each curve indicates  the $90\%$ confidence interval of the  curve.
Fig.~\ref{Fig1:Accuracy_N16_64_K20_40}-Left shows the test accuracy performance
 for $(N,K) = (64,20)$.
Our proposed JDU-BF outperforms other alternative schemes: it nearly attains the upper bound  under the Ideal FL after $40$ communication rounds and achieves an accuracy of $\sim 91\%$ at $\sim100$ rounds.
SDU-BF has a much slower model training convergence rate. After 100 rounds,
it only nearly reaches  $80\%$ test accuracy.
RBF exhibits the worst performance, where no training convergence is observed, and the accuracy is $\sim 10\%$ for all rounds.
This is because that RBF provides no beamforming gain, leading to highly suboptimal communication performance, which affects the learning performance.  Fig.~\ref{Fig1:Accuracy_N16_64_K20_40}-Middle shows the test accuracy for $(N,K) = (64,40)$.
We see that as the number of devices $K$ increases from $20$ to $40$, the learning performance and, thus, the test accuracy of  JDU-BF and SDU-BF improves. JDU-BF nearly attains the optimal performance after $30$ rounds, while SDU-BF approaches the upper bound  slowly
and is slightly worse than JDU-BF after 100 rounds.  The gain   comes from the improved uplink over-the-air aggregation as the result of (distributed)  transmit beamforming gain by more devices (\ie phase alignment via $a_{k,t}$).
In particular, the improvement of SDU-BF over $K$ is more noticeable.
RBF is still the worst among all methods, with the test accuracy remaining at $10\%$, as  it does not benefit from more devices since  no beamforming gain can be collected.

Fig.~\ref{Fig1:Accuracy_N16_64_K20_40}-Right shows the case for $(N,K) = (16,20)$, where $N<K$.
Compared with Fig.~\ref{Fig1:Accuracy_N16_64_K20_40}-Left,
 both JDU-BF and SDU-BF perform worse as $N$ reduces. This is expected due to reduced downlink and uplink beamforming gain with fewer antennas, impacting the overall learning performance of FL\ via wireless communication. Nonetheless, JDU-BF still nearly attains the upper bound after 100 rounds.
In summary,
our proposed JDU-BF is an effective communication scheme to facilitate FL  in a wireless system for achieving fast training convergence and high test accuracy.
\vspace*{-.7em}
\section{Conclusion}\vspace*{-.3em}
In this paper, we have formulated the downlink-uplink transmission process for  FL in a wireless system. We obtain
the global model update in each round, capturing the impact of transmitter/receiver processing, receiver noise, and local training on the model update. Aiming to optimize downlink-uplink beamforming to maximize the FL training performance, we have derived an upper bound on the expected global loss after $T$ rounds, and proposed  an efficient JDU-BF algorithm to minimize this upper bound.
JDU-BF is a low-complexity algorithm that uses the AO approach along with PGD to minimize the bound on the loss function per round.
Simulation results show that JDU-BF outperforms other alternative schemes and provides a near-optimal learning performance for wireless FL.

\appendices
\vspace*{-.7em}
\section{Proof of Proposition~\ref{thm:convergence}}\label{appC}\vspace*{-.3em}
\IEEEproof
Substitute $\rho_{k,t}=\frac{\alpha^{\text{ul}}_{k,t}}{\sum_{j=1}^{K}\alpha^{\text{ul}}_{j,t}}$ with
$\alpha^{\text{ul}}_{k,t}=\sqrt{p_{k,t}}|\hbf^{H}_{k,t}\wbf^{\text{ul}}_t|$ into \eqref{eq_A1_bound}\eqref{eq_A2t_lemma2}. We apply Lemmas~\ref{lemma1} and \ref{lemma2} to \eqref{theta_diff}. Let $M_t=\frac{\eta_tJ}{2Q_t}\big(5(1-Q_t)+4\sqrt{1-Q_t}-1\big)$ and  $R_t=H(\wbf^{\text{dl}}_t,  \wbf^{\text{ul}}_t,\pbf_t)+C_t$.
For $\frac{1}{10L}\le \eta_tJ<\frac{1}{2L}$, we have $Q_t>0$ and $5(1-Q_t)+4\sqrt{1-Q_t}-1>0$.
Thus, $M_t>0$ and $R_t>0$.
Then, after combining  Lemmas~\ref{lemma1} and \ref{lemma2} and \eqref{theta_diff}, we have
\vspace*{-.5em}
\begin{align}
&\mae[F(\thetabf_{t+1})]\!-\!F^{\star} \le\mae[F(\thetabf_t)]\!-\!F^{\star}\!+\!M_t\mae[\|\nabla F(\thetabf_t)\|^2]\!+\!R_t  \nn\\[-.1em]
& \stackrel{(a)}{\leq} \mae[F(\thetabf_t)]\!-\!F^{\star}\!+\!L^2M_t\mae[\left\|\thetabf_t - \thetabf^{\star}\right\|^2]\!+\!R_t \nn\\
& \stackrel{(b)}{\leq} \mae[F(\thetabf_t)]\!-\!F^{\star}\!+\!\frac{2L^2M_t}{\lambda}(\mathbb{E}[F(\thetabf_t)] - F^{\star})\!+\!R_t \label{eq_diff_upper_boundHC}
\end{align}
where $(a)$ and $(b)$ follow from \eqref{eq_global_local_equation} and  the $L$-smoothness and  $\lambda$-strong-convexity of $F_k(\cdot)$ in   Assumption~\ref{assump_smooth}, respectively.
Summing up both sides of \eqref{eq_diff_upper_boundHC} over $t\in\Tc$ and  rearranging the terms,
we have \eqref{eq_thm1}.
\endIEEEproof

\vspace*{-.5em}
\bibliographystyle{IEEEtran}
\bibliography{Refs}

\begin{thebibliography}{10}
\providecommand{\url}[1]{#1}
\csname url@samestyle\endcsname
\providecommand{\newblock}{\relax}
\providecommand{\bibinfo}[2]{#2}
\providecommand{\BIBentrySTDinterwordspacing}{\spaceskip=0pt\relax}
\providecommand{\BIBentryALTinterwordstretchfactor}{4}
\providecommand{\BIBentryALTinterwordspacing}{\spaceskip=\fontdimen2\font plus
\BIBentryALTinterwordstretchfactor\fontdimen3\font minus
  \fontdimen4\font\relax}
\providecommand{\BIBforeignlanguage}[2]{{%
\expandafter\ifx\csname l@#1\endcsname\relax
\typeout{** WARNING: IEEEtran.bst: No hyphenation pattern has been}%
\typeout{** loaded for the language `#1'. Using the pattern for}%
\typeout{** the default language instead.}%
\else
\language=\csname l@#1\endcsname
\fi
#2}}
\providecommand{\BIBdecl}{\relax}
\BIBdecl

\bibitem{Mcmahan&etal:2017}
B.~{McMahan}, E.~{Moore}, D.~{Ramage}, S.~{Hampson}, and B.~A.~Y. {Arcas},
  ``Communication-efficient learning of deep networks from decentralized
  data,'' in \emph{Proc. Int. Conf. Artif. Intell. Statist.}, Apr. 2017, pp.
  1273--1282.

\bibitem{Zhu&etal:2020}
G.~{Zhu}, D.~{Liu}, Y.~{Du}, C.~{You}, J.~{Zhang}, and K.~{Huang}, ``Toward an
  intelligent edge: {W}ireless communication meets machine learning,''
  \emph{IEEE Commun. Mag.}, vol.~58, no.~1, pp. 19--25, Jan. 2020.

\bibitem{Chen&etal:JSAC2012FLWN}
M.~{Chen}, D.~{Gündüz}, K.~{Huang}, W.~{Saad}, M.~{Bennis}, A.~V. {Feljan},
  and H.~V. {Poor}, ``Distributed learning in wireless networks: {R}ecent
  progress and future challenges,'' \emph{{IEEE} J. Sel. Areas Commun.},
  vol.~39, no.~12, pp. 3579--3605, Dec. 2021.

\bibitem{Yang&etal:TWC2020}
K.~{Yang}, T.~{Jiang}, Y.~{Shi}, and Z.~{Ding}, ``Federated learning via
  over-the-air computation,'' \emph{IEEE Trans. Wireless Commun.}, vol.~19,
  no.~3, pp. 2022--2035, Mar. 2020.

\bibitem{Zhu&etal:TWC2020}
G.~{Zhu}, Y.~{Wang}, and K.~{Huang}, ``Broadband analog aggregation for
  low-latency federated edge learning,'' \emph{IEEE Trans. Wireless Commun.},
  vol.~19, no.~1, pp. 491--506, Jan. 2020.

\bibitem{Zhang&Tao:TWC2021}
N.~{Zhang} and M.~{Tao}, ``Gradient statistics aware power control for
  over-the-air federated learning,'' \emph{IEEE Trans. Wireless Commun.},
  vol.~20, no.~8, pp. 5115--5128, Aug. 2021.

\bibitem{Sun&etal:JSAC2022}
Y.~{Sun}, S.~{Zhou}, Z.~{Niu}, and D.~{Gündüz}, ``Dynamic scheduling for
  over-the-air federated edge learning with energy constraints,'' \emph{{IEEE}
  J. Sel. Areas Commun.}, vol.~40, no.~1, pp. 227--242, Jan. 2022.

\bibitem{Fan&etal:TWC2022}
X.~{Fan}, Y.~{Wang}, Y.~{Huo}, and Z.~{Tian}, ``Joint optimization of
  communications and federated learning over the air,'' \emph{IEEE Trans.
  Wireless Commun.}, vol.~21, no.~6, pp. 4434--4449, Jun. 2022.

\bibitem{Amiri&etal:TWC2022}
M.~M. {Amiri}, D.~{Gündüz}, S.~R. {Kulkarni}, and H.~V. {Poor}, ``Convergence
  of federated learning over a noisy downlink,'' \emph{IEEE Trans. Wireless
  Commun.}, vol.~21, no.~3, pp. 1422--1437, Mar. 2022.

\bibitem{Wei&Shen}
X.~{Wei} and C.~{Shen}, ``Federated learning over noisy channels: {C}onvergence
  analysis and design examples,'' \emph{IEEE Trans. Cogn. Commun.}, vol.~8,
  no.~2, pp. 1253--1268, Jun. 2022.

\bibitem{Guo&etal:JSAC2022}
W.~{Guo}, R.~{Li}, C.~{Huang}, X.~{Qin}, K.~{Shen}, and W.~{Zhang}, ``Joint
  device selection and power control for wireless federated learning,''
  \emph{{IEEE} J. Sel. Areas Commun.}, vol.~40, no.~8, pp. 2395--2410, Aug.
  2022.

\bibitem{Wang&etal:JSAC2022b}
Z.~{Wang}, Y.~{Zhou}, Y.~{Shi}, and W.~{Zhuang}, ``Interference management for
  over-the-air federated learning in multi-cell wireless networks,''
  \emph{{IEEE} J. Sel. Areas Commun.}, vol.~40, no.~8, pp. 2361--2377, Aug.
  2022.

\bibitem{Levitin&Polyak:USSR1966}
E.~S. Levitin and B.~T. Polyak, ``Constrained minimization methods,''
  \emph{USSR Comput. Math. Math. Phys.}, vol.~6, no.~5, pp. 787--823, 1966.

\bibitem{Bubeck:2015convex}
S.~{Bubeck}, ``Convex optimization: {A}lgorithms and complexity,'' \emph{Found.
  Trends Mach. Learn.}, vol.~8, no. 3-4, pp. 231--357, 2015.

\bibitem{Ghadimi&etal:Math2016}
S.~Ghadimi, G.~Lan, and H.~Zhang, ``Mini-batch stochastic approximation methods
  for nonconvex stochastic composite optimization,'' \emph{Math. Program.},
  vol. 155, no. 1-2, pp. 267--305, 2016.

\bibitem{Zhang&etal:WCL2022}
C.~{Zhang}, M.~{Dong}, and B.~{Liang}, ``Fast first-order algorithm for
  large-scale max-min fair multi-group multicast beamforming,'' \emph{IEEE
  Wireless Commun. Lett.}, vol.~11, no.~8, pp. 1560--1564, Aug. 2022.

\bibitem{Dong&Wang:TSP2020}
M.~{Dong} and Q.~{Wang}, ``Multi-group multicast beamforming: Optimal structure
  and efficient algorithms,'' \emph{{IEEE} Trans. Signal Process.}, vol.~68,
  pp. 3738--3753, May 2020.

\bibitem{Lecun&etal:MNIST}
Y.~{LeCun}, C.~{Cortes}, and C.~{Burges}. {The MNIST Database of Handwritten
  Digits}. 1998. [Online]. Available: http://yann.lecun.com/exdb/mnist/.

\end{thebibliography}

\end{document}